\documentclass[aps,pre,preprint,amsmath,amssymb,groupedaddress,showpacs]{revtex4-1}
\usepackage{graphicx}
\usepackage{dcolumn}
\begin{document}

% \preprint{}

\title{Energy landscape theory for cotranslational protein folding}


\author{David S. Tourigny}
\email[]{davidt@mrc-lmb.cam.ac.uk}
%\thanks{}
\affiliation{MRC Laboratory of Molecular Biology, Cambridge CB2 0QH, UK}

%\no affiliation

\date{\today}

\begin{abstract}
Energy landscape theory describes how a full-length protein can attain its native fold after sampling only a tiny fraction of all possible structures. Although protein folding is now understood to be concomitant with synthesis on the ribosome there have been few attempts to modify energy landscape theory by accounting for cotranslational folding. This paper introduces a model for cotranslational folding that leads to a natural definition of a nested energy landscape. By applying concepts drawn from submanifold differential geometry the dynamics of protein folding on the ribosome can be explored in a quantitative manner and conditions on the nested potential energy landscapes for a good cotranslational folder are obtained. A generalisation of diffusion rate theory using van Kampen's technique of composite stochastic processes is then used to account for entropic contributions and the effects of variable translation rates on cotranslational folding. This stochastic approach agrees well with experimental results and Hamiltionian formalism in the deterministic limit. 
\end{abstract}

%\keywords{}

\maketitle

%\section{}
%\subsection{}
%\subsubsection{}

\section{Introduction}
A fundamental problem in molecular biology is explaining how the three-dimensional structure of a protein is encoded within its amino acid sequence. Inside the cell, proteins are synthesised on the ribosome by sequential addition of residues to an elongating polypeptide chain during a process called translation \cite{Schmeing09}. Translation accounts for the conversion of genetic information to the primary sequence of a protein, but knowledge of how the molecule then folds into a functional state is central to our understanding of the natural world. Energy landscape theory provides a mechanism whereby the existence of intermediate structures, each associated with a free energy cost, enable the folding pathway of a protein to be mapped on a multidimensional potential energy landscape. Assuming the global shape of the energy landscape for a good folder resembles a funnel means that only a small fraction of all possible structures need to be sampled before the protein attains its native fold \cite{Leopold92,Bryngelson95,Onuchic95}.    

Nascent proteins can begin to fold whilst they are still bound to a translating ribosome \cite{Fedorov97,Cabrita10}. During cotranslational folding the conformational space available to a protein increases incrementally with addition of residues to the polypeptide chain. This can enhance folding yields \cite{Ugrinov10}, provide an additional level of quality control \cite{Netzer97,Nicola99}, and allow access to folding pathways different from those available to a full-length protein \cite{Clark01,Elcock06}. Studies have revealed a relationship between folding timescales and dwell times before amino acid addition that can be modulated by synonymous codon mutations and controlled by the translational apparatus \cite{Komar99,Tsai08,Zhang09}. Some research groups have also developed a theoretical understanding of how protein folding is affected by varying translation rates \cite{O'Brien12,Ciryam13}, but so far there has not been a satisfactory attempt to modify energy landscape theory by accounting for cotranslational folding. 

A suitable energy landscape theory would extract some general properties of cotranslational folders from the complex nature of the system. A key question that may be answered using energy landscape theory is: what distinguishes cotranslational folders from arbitrary polymers or proteins that can only fold once translation is over? Moreover, how do the effects of synonymous mutations and variable translation rates manifest themselves on folding pathways and time scales? The remainder of this paper will address these questions. In Section II, the concept of a nested energy landscape is introduced for a geometrisation of Hamiltonian dynamics. This construction is extended to free energy landscapes in Section III, where the diffusion rate theory of Bryngelson and Wolynes is generalised to cotranslational protein folding. Some numerical calculations are presented in Section IV where there is an emphasis on the role of translation rate in determining folding rates, yields, and pathways. This is cemented by some concluding remarks in the final section.   

\section{Geometry of the potential energy landscape}
A geometric approach should reveal which energy landscape shapes can accommodate variable translation rates. This is important for identifying good cotranslational folders and providing some upper limit to the rate of translation based on their energy landscapes. In the past, pseudo-Riemannian geometry has been successfully applied to the folding of full-length proteins \cite{Mazzoni06,Mazzoni08}. Although the results follow a similar theme, this section develops a new theory specific to cotranslational folding.

\subsection{Preliminaries of nested energy landscape theory}
An enlarged ($N+2$)-dimensional space with coordinates $r^0,r^1,r^2,...,r^N, r^{N+1}$ can be used to describe the configuration space of a polypeptide of length $n$ \cite{Mazzoni06,Mazzoni08}. The potential energy as a function of the reaction coordinate $\mathbf{r}=(r^1,r^2,...,r^N)$ is denoted by $V_n(\mathbf{r})$. In keeping with the assumptions of  \cite{O'Brien12,Ciryam13}, the transition from nascent chain length $(n-1)$ to $n$ is instantaneous relative to the times that the ribosome spends at either of these chain lengths and so $\mathbf{r}$ does not change upon the addition of a new amino acid. The configuration spaces of the $n$th and $(n-1)$th states can be represented as manifolds $M$ and $\bar{M}$ respectively, and it is appropriate to endow $M$ with an Eisenhart metric whose arc length is 
\begin{equation} ds^2= \delta_{ij}dr^idr^j -2V_n(\mathbf{r})(dr^0)^2+2dr^0dr^{N+1} \ , \end{equation}
where the indices $i,j$ run from $1$ to $N$ \cite{Eisenhart28}. Projected geodesics of this Eisenhart metric are natural motions of the Hamiltonian system and therefore the folding trajectories of a protein in the $n$th state. These trajectories, parameterised by the time coordinate $r^0=t$, are obtained by taking $ds^2=dt^2$ on the physical geodesics and imposing the integral condition
\begin{equation} r^{N+1}= \frac{1}{2}t +c_0 - \int_0^t\  [\delta_{ij} \dot{r}^i\dot{r}^j  -V_n(\mathbf{r})]  \ dt \end{equation}     
on the additional coordinate $r^{N+1}$. Here $c_0$ is some arbitrary real constant.

Supposing there to be a differentiable isometric immersion $f :\bar{M} \to M$, for each  $p \in \bar{M}$ there exists a neighbourhood of $\bar{M}$ whose image is a submanifold of $M$. The immersion $f$ is used to define what is meant by saying that the $(n-1)$th energy landscape is a nested energy landscape of the $n$th. At $p$ the decomposition
\begin{equation} T_pM=T_p\bar{M} \oplus (T_p\bar{M})^{\perp} \ ,\end{equation} 
states that the tangent vector space $T_pM$ can be decomposed into a direct sum of the tangent space $T_p\bar{M}$ and its orthogonal complement $(T_p\bar{M})^{\perp}$. As a consequence, $\bar{M}$ inherits a metric and affine connection $\bar{\nabla}$ from the Eisenhart metric of $M$. For $X$ and $Y$ vector fields on $\bar{M}$ extended to $M$ it can be shown that $\bar{\nabla}_X Y$ is equal to the component of $\nabla_X Y$  tangential to $\bar{M}$, where $\nabla$ is the affine connection on $M$ \cite{doCarmo92}. The difference $\nabla - \bar{\nabla}$ uniquely defines the mapping $H: T_p\bar{M} \times T_p\bar{M} \to (T_p\bar{M})^{\perp}$ and the \emph{shape operator} $S_Z$:

\begin{equation}  \langle S_Z(X), Y \rangle = \langle H(X,Y), Z \rangle \end{equation}
along the direction $Z \in (T_p\bar{M})^{\perp}$. The eigenvalues of $S_Z$ are called the \emph{principle curvatures} of $\bar{M}$, and are a measure of how much the manifold $\bar{M}$ bends towards $Z$ at the point $p$. These concepts will be used later in this section. 

\subsection{Focal points and convergence of folding pathways}
The time of addition of the $n$th amino acid is denoted $t_n$ so that $t_{n+1}=t_n + \tau_{A,n+1}$ when $ \tau_{A,n+1}$ is the dwell time at the $n$th codon. The point $\gamma_0 \in \bar{M} \subset M$ is the point of the immersed space $\bar{M}$ where the $n$th amino acid was added to the nascent protein. That particular folding trajectory is then no longer constrained to $\bar{M}$, but can continue as a geodesic $\gamma: [t_n,t_{n+1}] \to M$ with initial tangent vector $\dot{\gamma}_0 \in (T_{\gamma_0}\bar{M})^{\perp}$ that guarantees $\gamma$ will leave $\bar{M}$. 

To achieve a stable fold on the ribosome before addition of the $(n+1)th$ amino acid, any similarly constructed trajectory $\sigma$, resulting from a delay in addition of the $n$th amino acid and emanating elsewhere on $\bar{M}$, must converge with $\gamma$ over the given time interval $\tau_{A,n+1}$. The distance to $\sigma$ from any point along $\gamma$ is measured by the Jacobi vector field $J\in T_{\gamma} M$, which is everywhere orthogonal to the tangent vector field $\dot{\gamma} \in T_{\gamma} M$ and must satisfy the Jacobi equation
\begin{equation} \nabla_{\dot{\gamma}} \nabla_{\dot{\gamma}} J = R(\dot{\gamma},J) \dot{\gamma} \ . \end{equation}
Here $R$ is the Riemann curvature tensor on $M$, whose only non-vanishing components in the chosen coordinate chart are $R_{0i0j}=\partial_i \partial_j V_n(\mathbf{r})$. A small $\| J \|$ implies stability along $\gamma$, whereas large $\| J \|$ is indicative of chaotic behaviour \cite{Casetti00}; a point along $\gamma$ at which $J$ vanishes and trajectories converge to a common fold is called a \emph{focal point} of $\bar{M}$. 

Variable translation rates mean that to achieve a stable fold at chain length $n$, the nested energy landscape of a good cotranslational folder must be sculpted in a way that ensures convergence of perturbed folding trajectories. For this to be one of the criteria of good folders implies that their energy landscapes can be distinguished geometrically from those of general polymers. The time taken to reach the first focal point of $\bar{M}$ along $\gamma$ is completely determined by the curvatures of $M$ and $\bar{M}$, and so it is possible to deduce the geometric features of suitable landscapes.    

\subsection{Conditions for a focal point over the interval $\tau_{A,n+1}$}      
To attain a stable fold at chain length $n$ a focal point must be reached before addition of the $(n+1)$th amino acid. From Proposition 10.35 in \cite{O'Neill83} it is possible to derive conditions on $\bar{M}$ and $M$ that guarantee a focal point of $\bar{M}$ over  $(t_n,t_{n+1}]$. The quadratic form defined by 
\begin{equation} h_{\dot{\gamma}_0}(X)=\langle H(X,X), \dot{\gamma}_0 \rangle \end{equation}
 for some unit vector $X \in T_{\gamma_0} \bar{M}$ is called the \emph{second fundamental form} of $\bar{M}$ at $\gamma_0$ along the direction $\dot{\gamma}_0$ \cite{doCarmo92}. It turns out there is always a focal point along $\gamma$ before addition of the $(n+1)$th amino acid provided that $h_{\dot{\gamma}_0}(X) \geq 1/\tau_{A,n+1}$ and the \emph{sectional curvatures} of all two-planes containing $\dot{\gamma}$ are positive semidefinite. This is a powerful result, but the unfortunate dependence of the conditions on arbitrary choices of vectors and two-planes makes it difficult to grasp the requirement on nested energy landscapes.
 
It would be preferable to derive a simpler relationship between curvatures and the distance to the first focal point of $\bar{M}$. This can be achieved by introducing a new construction on $M$, but with a cost of ambiguity added to the location of the focal point. It is always possible to pick a hypersurface $P\subset M$ through $\gamma_0$ orthogonal to $\dot{\gamma}_0$ so that at $\gamma_0$ the shape operator of $P$ agrees with $S_{\dot{\gamma}_0}$. From Warner \cite{Waner66}, the first focal point of any such $P$ occurs at least as soon as the first focal point of $\bar{M}$ and so the best choice of $P$ is the hypersurface whose first focal point occurs furthest along $\gamma$. Adapting the proof of Proposition 10.37 in \cite{O'Neill83}, the following conditions must be satisfied if a focal point of $\bar{M}$ is to occur over $(t_n,t_{n+1}]$. Provided that
\begin{equation} \frac{1}{N+1}\  \mbox{trace}(S_{\dot{\gamma}_0}) \geq \frac{1}{\tau_{A,n+1}}   \label{qg1} \ ,\end{equation}
and 
\begin{equation} \frac{1}{N+1} \ \mbox{Ric}(\dot{\gamma},\dot{\gamma}) = \frac{1}{N+1} \ \Delta V_n(\mathbf{r}) \geq 0   \label{qg2} \ , \end{equation}
there can exist a focal point of $\bar{M}$ on $\gamma$ over the interval $(t_n,t_{n+1}]$. The operator $\mbox{Ric}: T_pM \times T_pM \to \mathbb{R}$ appearing in (\ref{qg2}) is the Ricci tensor on $M$ whose only non-vanishing components in the chosen coordinate chart are $R_{00}=\Delta V_n(\mathbf{r})$. By combining Corollary 2.3 in Warner \cite{Waner66} with a theorem of Myers \cite{Myers41} it can also be shown that convergence before $t_{n+1}$ is guaranteed whenever 
\begin{equation}   \frac{1}{N+1} \ \Delta V_n(\mathbf{r})  \geq \frac{\pi^2}{\mbox{Length} \{ \gamma(t_n,t_{n+1}) \}}  \ , \label{qg3} \end{equation} 
no matter how perturbed folding trajectories originate from the $(n-1)$th energy landscape. Whilst unlikely to be satisfied by all potential landscapes, the much stronger condition (\ref{qg3}) serves to guarantee that a stable fold will be attained at chain length $n$. 

\begin{figure}
\centering
\includegraphics[width=0.6 \textwidth]{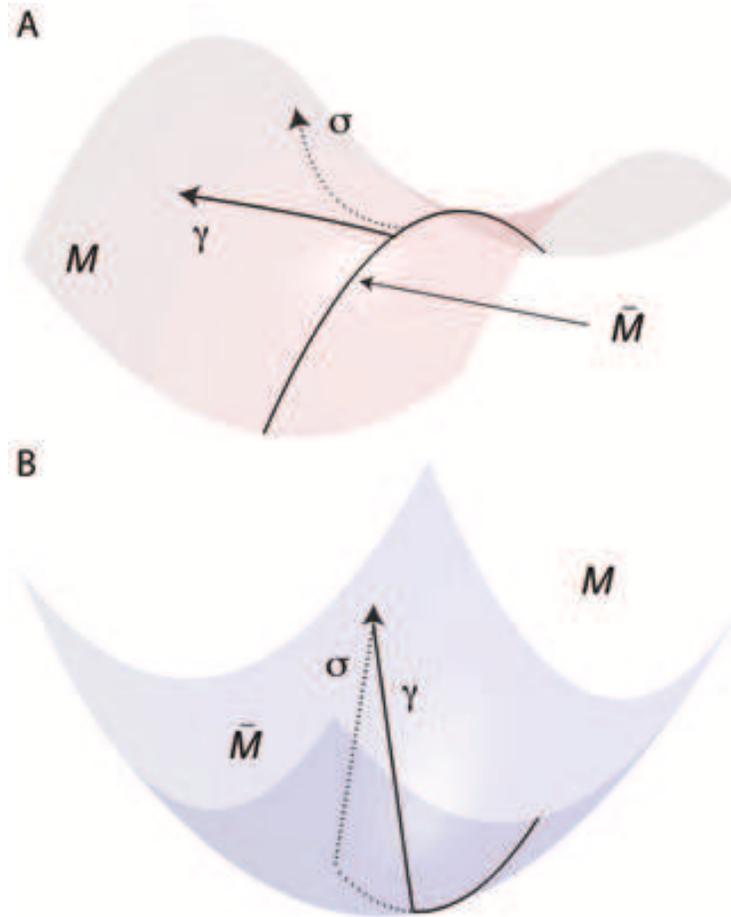}
\caption{A simple scheme for an intuitive grasp of conditions on the nested energy landscapes of a good cotranslational folder. (A) Perturbed folding trajectories leaving a nested energy landscape will be unable to converge when entering a region of non-positive curvature. (B) Folding trajectories must leave nested energy landscapes at a point where curvature is sufficient for ensuring convergence over the available time interval.}
\label{Fig1}
\end{figure}

The minimal requirement for perturbed trajectories to converge at a certain chain length is that amino acids be added where the average curvature of $V_{n-1}(\mathbf{r})$ is proportional to the rate of translation and $\Delta V_n(\mathbf{r})$ is positive definite along the folding pathway (Fig. \ref{Fig1}). Similarly to the folding of full-length proteins \cite{Mazzoni06,Mazzoni08}, it would appear that the average curvatures of nested energy landscapes are enough to distinguish a good cotranslational folder from arbitrary homopolymers. Moreover, average curvatures increase with the probability of attaining a stable fold at longer chain lengths. 

\section{Diffusive dynamics on the free energy landscape}
For a complete description of a folding pathway the change in entropy must also be accounted for as the reaction coordinate moves around the energy landscape. It is therefore the profile of a free energy landscape that determines folding rates and pathways, and superimposed on this are stochastic contributions from the reaction process. The purpose of this section is to develop a statistical theory of cotranslational folding that generalises deterministic behaviour through the mean folding time.   

\subsection{Cotranslational folding as a composite stochastic process} 
The nested energy landscape scheme outlined in Section II can be adapted by associating with each state $n$ a general operator $\mathbb{A}_n$ (the $n$th propagator) that describes the concerted change in both energy and entropy as a function of conformational changes in the nascent chain. Since the length of time spent at each state is stochastic by nature, the jump from state $n$ to $(n+1)$ occurs with a transition probability $\lambda_{n,n+1}(\tau)$ per unit time, which is a function of the time, $\tau$, sojourned in state $n$. The probability that the protein chain remains at length $n$ after time $t_n+\tau$ is defined by

\begin{equation} u_n(\tau) = \mbox{exp} \left[- \int_{t_n}^{t_n + \tau}  \lambda_{n,n+1}(\tau') d \tau' \right] \ , \end{equation}
and the quantity $v_{n,n+1} (\tau) = u_n (\tau) \lambda_{n,n+1}(\tau)$ is the probability that the $(n+1)$th amino acid is added at a time $t_n+ \tau$. The probabilities $v_{n,n+1}(\tau)$ have yet to be experimentally determined, but if assumed to be exponential functions with mean $\langle \tau_{A,n+1} \rangle$, the $\lambda_{n,n+1}$ become independent of $\tau$ and the process becomes Markovian. This is equivalent to the approximation made in \cite{O'Brien12}, but it is not made here.

In general, the probability at any time $t$ for finding the protein to be of length $n$ and in conformation $\mathbf{r}$ is given by the conditional probability $P_n(\mathbf{r},t|\mathbf{r}_0,t=0)$ that describes a non-Markovian process beginning in conformation $\mathbf{r}_0$ at time $t=0$. Assuming, as in Section II, that the position of $\mathbf{r}$ does not change upon the transition from the $(n-1)$th to the $n$th state, and working in the dimension of the $n$th configuration space, this probability density describes a composite stochastic process \cite{VanKampen79}. The marginal probability distribution $P(\mathbf{r},t|\mathbf{r}_0,t=0)$ corresponds to the protein being in conformation $\mathbf{r}$ at time $t$ regardless of chain length

\begin{equation}  P(\mathbf{r},t|\mathbf{r}_0,t=0) = \sum_n P_n(\mathbf{r},t|\mathbf{r}_0,t=0) \ . \end{equation}

The mean first passage time for reaching the correctly folded conformation ($\textbf{r}_F$) from any given starting conformation $\textbf{r}_0$ at time $t=0$, is given by

\begin{equation} \tau_F = - \frac{\partial \hat{F}(\textbf{r}_0,s)}{\partial s} \bigg|_{s=0} \ , \end{equation}
where, using an extension of the renewal equation \cite{VanKampen93}, 

\begin{equation} \hat{F}(\textbf{r}_0,s) = \frac{\hat{P}(\textbf{r}_F,s|\textbf{r}_0,t=0)}{\hat{P}(\textbf{r}_0,s|\textbf{r}_0,t=0)} \ ,\end{equation}   
where $\hat{P}(\textbf{r},s|\textbf{r}_0,t=0)$ is the Laplace transform of (12). From \cite{VanKampen79} it is possible to derive 

\begin{equation} \hat{P}(\textbf{r},s|\textbf{r}_0,t=0) = \sum_n   \left( \frac{}{}[\hat{U}(s)]_{n,n}-[\hat{U}(s)\cdot \hat{V}(s)]_{n,n+1} \right) P_n(\mathbf{r}_0,t=0) \ , \end{equation}
where $\hat{U}(s)$ and $\hat{V}(s)$ are the Laplace transforms of matrix operators whose only non-zero components are $[U(\tau)]_{n,n}=u_n(\tau) \exp(\tau \mathbb{A}_n)$ and $[V(\tau)]_{n,n+1} = v_{n,n+1}(\tau) \exp(\tau \mathbb{A}_n)$, respectively. These equations are the central result of this section: an expression for the mean folding time given an arbitrary set of dwell time distributions and propagators. Mean folding times are important since they play a role in defining a folding pathway precisely \cite{Wang96}.  

\subsection{The effective landscape approximation} 
The form of the propagators $\mathbb{A}_n$ must be chosen appropriately to obtain a reasonable approximation for $\tau_F$. Bryngelson and Wolynes \cite{BW87,BW89} first suggested that protein folding can be described by diffusion of a reaction coordinate depicting distance from the folded configuration on a one-dimensional free energy landscape. Since diffusion rate theory agrees extremely well with the results of folding simulations \cite{Socci96} it makes sense to take 

\begin{equation} \mathbb{A}_n = \frac{\partial}{\partial r} \left[D_n(r) \frac{\partial}{\partial r} +  D_n(r) \frac{\partial \beta G_n(r)}{\partial r}  \right] \ , \end{equation}
where $\beta$ is the inverse of temperature ($T$) and $G_n(r)$ is the $n$th free energy landscape. The diffusion coefficient $D_n(r)$ reflects the ruggedness of the $n$th landscape in the proximity of the glass transition temperature $T_{g,n}$ for that landscape. 

Good folding sequences of length $n$ will have a folding temperature $T_{F,n}>T_{g,n}$. At folding temperatures $T_{F,n}>2T_{g,n}$, using the law of typical glasses $D_n(r)$ is given by

\begin{equation} D_n(r) = D_0\ \mbox{exp} [-\beta^2 \Delta E_n^2(r)] \ , \end{equation}
where $\Delta E_n^2(r)$ is the local mean fluctuation in energy and $D_0$ is the diffusion coefficient on a flat landscape. For folding temperatures $2T_{g,n}(r) > T_{F,n} > T_{g,n}(r)$ a more suitable approximation is

\begin{equation}  D_n(r) = D_0\ \mbox{exp} [-S_n^{\star}(r) +(\beta_{g,n}(r)-\beta)^2 \Delta E_n^2(r)] \ , \end{equation}
where $S_n^{\star}(r)$ is the configurational entropy of $r$ in the $n$th state. At $T_{g,n}$ the diffusion coefficient decays exponentially from $D_0$ by a factor of the total number of configuration states. The diffusion coefficient must therefore increase with chain length $n$ to prevent trapping in local minima and a Levinthal paradox at temperatures near $T_{g,n}$.        

It is well understood that slow translation rates will afford a protein more time to fold on the ribosome \cite{Zhang09,Purvis87}. In these cases cotranslational folding becomes a quasi-equilibrium process that can be approximated by its thermodynamically determined value. The folding process of a protein domain is slower at the surface of the ribosome however \cite{Kaiser11}, and so it has been estimated that kinetic effects dominate folding for more than 20 \% of the \emph{Escherichia coli} cytosolic proteome \cite{O'Brien12,Ciryam13}. In situations where the average dwell time $\langle \tau_{A,n+1} \rangle $ is short compared to $\tau_F$ the marginal distribution $P(r,t)=\sum_n P_n(r,t)$ satisfies an effective random walk governed by the master equation

\begin{equation} \frac{\partial P(r,t) }{\partial t} = \mathbb{A}_e\ P(r,t) \ , \end{equation}  
with an effective propagator given by $ \mathbb{A}_e = \sum_n \alpha_n  \mathbb{A}_n$. Here

\begin{equation} \alpha_n = \frac{\langle \tau_{A,n+1} \rangle \zeta_n}{\sum_n \langle \tau_{A,n+1} \rangle \zeta_n} \ , \end{equation}
and the $\zeta_n$ are the components of the right eigenvector of the array $\hat{v}_{n,n+1}(s=0)$ satisfying $\sum_n \zeta_n =1$ \cite{VanKampen79} . It is important to note that this result is not restricted to the one-dimensional case, but taking $\mathbb{A}_n$ to be the Bryngelson and Wolynes operator allows the mean folding time to be written as the double integral

\begin{equation} \tau_F = \int_{r_0}^{r_F} dr \int_0^r dr' \ \frac{\mbox{exp}[\beta G_e(r)-\beta G_e(r')]}{D_e(r)} \ .  \label{qg4} \end{equation}
$G_e(r)$ and $D_e(r)$ are, respectively, the \emph{effective free energy landscape} $G_e(r) = \sum_n \alpha_n G_n(r)$ and the \emph{effective diffusion coefficient} $D_e(r) = \sum_n \alpha_n D_n(r)$.  

When $G_e(r)$ is assumed double-welled with a barrier peak at the point $r_t$, the integral in (\ref{qg4}) can be approximated by 

\begin{equation} \tau_F \approx \left(\frac{2 \pi}{\beta}\right)^{1/2} \frac{1}{D_0 \omega_0 \bar{\omega}_F} \mbox{exp} [\beta \bar{G}_e(r_t) - \beta G_e(r_0)] \ , \label{gq7} \end{equation}
where

\begin{equation} \bar{G}_e(r) = G_e(r) - T \mbox{log} \left[ \frac{D_e(r)}{D_0} \right] \ , \end{equation}
and $\omega_0$ and $\bar{\omega}_F$ are the curvatures at $r_0$ and the top of the barrier respectively \cite{Socci96}. These expressions will be used in Section IV to predict the effect of variable translation rates on folding time scales for a protein on the ribosome. 

\subsection{Relation to Hamiltonian theory}
The effective landscapes and diffusion coefficients represent weighted averages of $G_n(r)$ and $D_n(r)$ over a subset of $n=1,2,...$, where the contribution from the $n$th landscape is proportional to the average amount of time spent at that state. The largest contributions to the shape and height of $G_e(r)$ are from the nested landscapes $n$ for which $\langle \tau_{A,n+1}\rangle$ is largest and the reaction coordinate spends most of its time. This dictates the size and position of the free energy peak that must be traversed by the protein in addition to the amplitude of the effective diffusion coefficient along the folding pathway. Considering each nested landscape $n$ individually, the condition for achieving a stable fold before addition of the $(n+1)$th amino acid is 

\begin{equation}  \left(\frac{\beta}{2 \pi}\right)^{1/2}D_0 \omega_0 \bar{\omega}_F \mbox{exp} [- \beta \bar{G}_n(r_t) + \beta G_n(r_0)]  \gtrsim \frac{1}{  \langle \tau_{A,n+1} \rangle} \ ,\label{qg5} \end{equation}
where $r_t,r_0$ and $\omega_0,\bar{\omega}_F$ are particular to the $n$th landscape. From the set of generic polymers with fixed $D_n(r_t)$, $G_n(r_t)$ and $G_n(r_b)$ for a given series of dwell times, it is only those sequences with sufficiently large curvatures $\omega_0$ and $\bar{\omega}_F$ for which this inequality holds, distinguishing protein-like sequences from arbitrary heteropolymers. It helps to think of the free energy landscape $G_n(r)$ as an averaged projection of connected minima of the nested potential energy landscape \cite{Wales06}. More precisely, there is an obvious correspondence between $\omega_0$ and $\mbox{trace}(S_{\dot{\gamma}_0})/(N+1)$ in condition (\ref{qg1}) from Section II. Likewise, $\bar{\omega}_F$ represents a measured average of $\Delta V_n(\mathbf{r})$ along the folding pathway which, in consistency with full-length simulations \cite{Mazzoni08}, is expected to increase with the fraction of native contacts $r$ as the protein approaches the folded state. This reestablishes the connection with conditions  (\ref{qg2}) and (\ref{qg3}) from Section II. 

\section{Cotranslational folding of a real protein}
An estimate of mean folding time will help to define each cotranslational folding pathway precisely. This allows the effects of variable translation rates on folding rates and yields to be described in relation to the new energy landscape theory. Protein G is a single-domain protein of $56$ amino acids for which coarse-grain cotranslational folding simulations have proven consistent with a range of experimental results \cite{O'Brien10,O'Brien11,O'Brien12}. This section will compare folding time scales and derive physically meaningful free energy landscapes from the results of these simulations.  

\subsection{The influence of synonymous mutations on mean folding time}
A quicker overall translation rate has been shown to reduce the probability of attaining a stable fold on the ribosome. Specifically, the effect of variable translation rates over the last 10 codons of a 35 amino acid extension of protein G has been investigated \cite{O'Brien12}. Decreasing the translation time by an order of magnitude per codon decreases the probability that a stable fold is achieved on the ribosome at a particular chain length. However, this decrease can be partially recovered by introducing a synonymous mutation with a slower translation at a single codon position.  

\begin{figure}
\centering
\includegraphics[width=0.9 \textwidth]{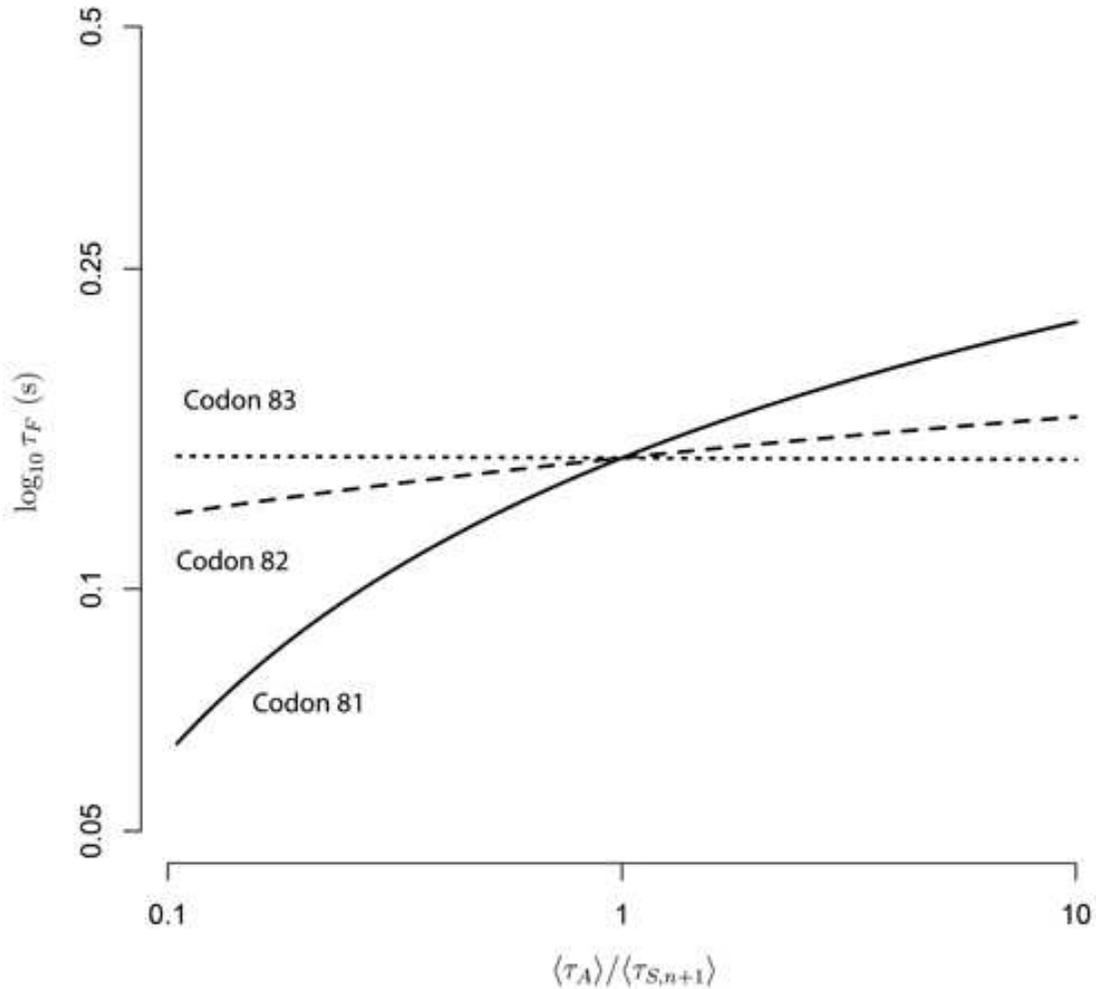}
\caption{Mean cotranslational folding time $\tau_F$ plotted as a function of synonymous mutation type at codon positions. Fast-to-slow synonymous mutations have a small $\langle \tau_A \rangle / \langle \tau_{S,n+1} \rangle$ ratio whereas the converse is true for slow-to-fast mutations. Only three codon positions are displayed for reasons of clarity. }
\label{Fig2}
\end{figure}

Setting $\mathbf{r}_0$ as the conformation of the nascent chain at time $t_{81}=0$, a suitable approximation for the mean first passage time to $\mathbf{r}_F$ using the model from Section III is 

\begin{equation} \tau_F \approx \sum_{n=81}^{91} \frac{\langle \tau_{A,n+1} \rangle  \tau_{F,n} }{11\sum_n \langle \tau_{A,n+1} \rangle} \ , \label{qg6} \end{equation}
where the $\tau_{F,n}$ are those reported in \cite{O'Brien12,Ciryam13}. These are the times it takes a protein arrested in an unfolded state at each chain length $n$ to reach a folded conformation, which can now be measured using an experimental setup on the ribosome \cite{Kaiser11}. Equation (\ref{qg6}) expresses the mean folding time as a function of the average translation speed for each codon. To simulate the effect of synonymous mutations on cotranslational folding $\tau_F$ can be calculated for every instance that codon $n$ is replaced by a codon translated at an average rate of $1/\langle \tau_{S,n+1} \rangle$. Some results are displayed in Fig. \ref{Fig2} where $\tau_F$ is plotted as a function of the ratio $\langle \tau_A \rangle / \langle \tau_{S,n+1} \rangle$ and $\langle \tau_A \rangle$ is the average translation time scale of the other 9 codons. 

Fig. \ref{Fig2} confirms a positional dependence on the effect any particular type of synonymous mutation has on cotranslational folding under kinetic control. When a mutation that decreases overall translation time is made at an earlier codon there is a significant increase in the mean folding time. Conversely, the same mutation at a later codon has little effect on the overall folding speed. This is rationalised by noting that at earlier times there is a lower probability for the protein to be in a stable fold and addition of the next amino acid contributes to instability. Geometrically, perturbed folding trajectories have not converged over the interval that the synonymous mutation has provided whereas a stable fold has already been attained by the time later codons are translated. For the same reason, fast-to-slow mutations at later positions also have little effect whilst at earlier positions these substitutions afford the protein greater time to attain a stable fold and improve the mean folding time. These results agree well experimental evidence for synonymous mutations having a pronounced effect on folding yields and efficiency \cite{Zhang09}.                     

\subsection{Derived free energy landscapes}
It only remains to confirm the existence of effective free energy landscapes and diffusion coefficients that enable the folding time scales predicted by equation (\ref{qg6}) to be described by a sensible set of parameters. A scheme implemented in C++ with the conformational entropy $S_n^{\star}(r)$ defined by Bryngelson and Wolynes \cite{BW89} as 

\begin{equation}  S_n^{\star}(r) = n \left[ -r\log r -(1-r)\log \left(\frac{1-r}{10}\right) \right] \ , \end{equation}
was used to search for parameters of the effective energy landscape. As an example, equation (\ref{qg6}) predicts $\tau_F$ to fall from $1.4 \times 10^{-1}$s to $6.4 \times 10^{-2}$s when a fast-to-slow synonymous mutation is made at codon 81. The 35 residues of the linker protein simply serve as an anchor to the ribosome exit tunnel and so only the 56 residues belonging to protein G are largely contributing to the effective free energy landscape. The functional forms of $G_e(r)$ and $D_e(r)$ were taken from \cite{BW89} and parameters were estimated from simulation data with $D_0 \approx 10^{9}\mbox{s}^{-1}$.     

\begin{figure}
\centering
\includegraphics[width=0.9 \textwidth]{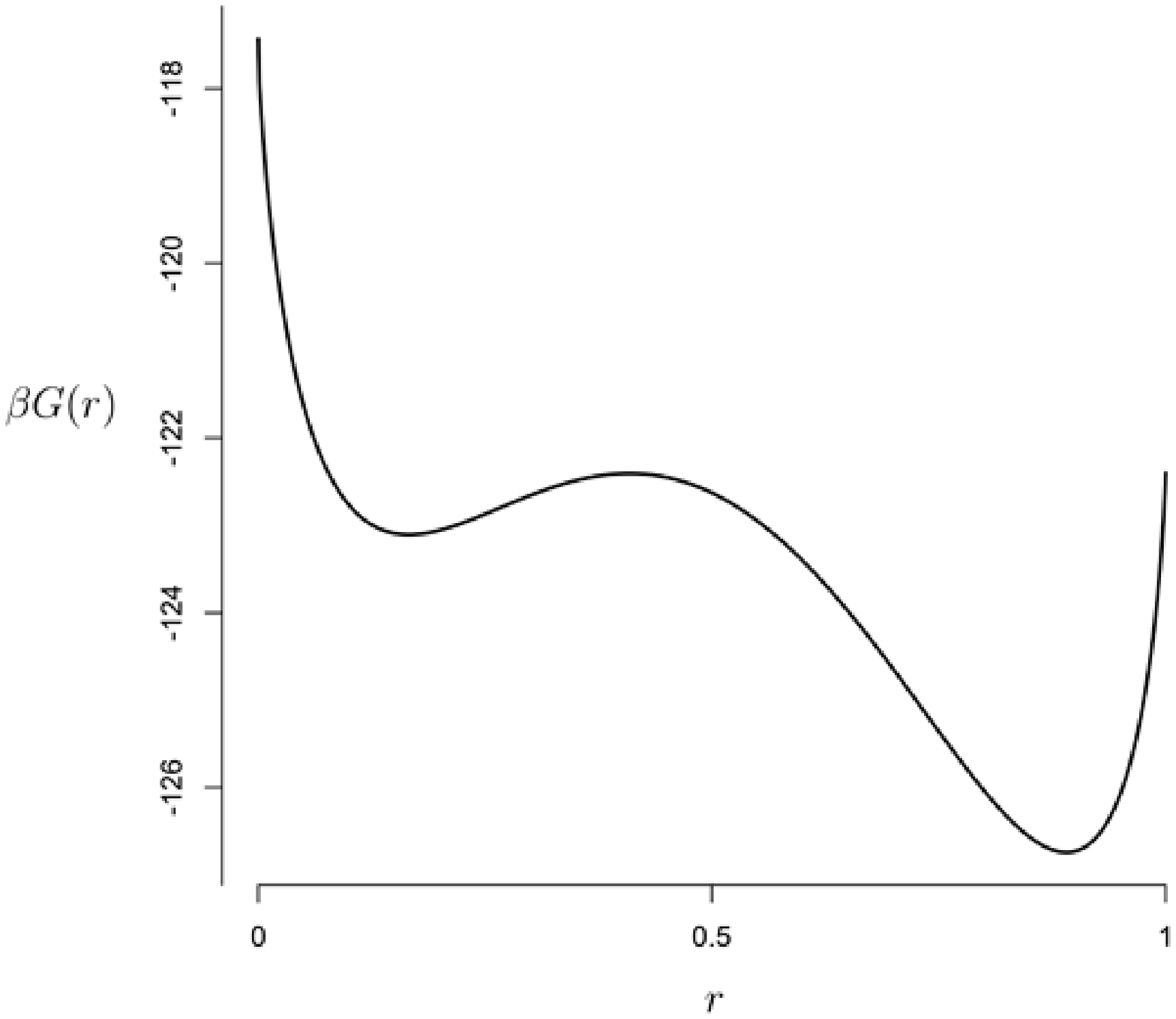}
\caption{The free energy landscape $G^{51}(r)$ as a function of native contacts.}
\label{Fig3}
\end{figure}

A double-welled landscape whose values are within realistic limits for a protein with $\sum_{n=46}^{56} \alpha_n n=51$ residues is plotted in Fig. \ref{Fig3} and given by

\begin{equation} G^{51}(r) = \mbox{8.16}Tr-\mbox{130.56}Tr^2+\mbox{51.0}Tr \log r + \mbox{51.0}T (1-r) \log \left(\frac{1-r}{10} \right) \ . \end{equation}
An appropriate diffusion coefficient is 
\begin{equation} D^{51}(r) = D_0 \exp [ - \beta^2 \mbox{4.53} (1-r) +\beta \mbox{0.51} (1-r^2) ] \ .\end{equation}
For $\sum_{n=46}^{56} \alpha_n n=46.5$ residues
 
\begin{equation} G^{46.5}(r) = \mbox{7.36}Tr-\mbox{117.76}Tr^2+\mbox{46.5}Tr \log r + \mbox{46.5}T (1-r) \log \left(\frac{1-r}{10} \right) \ , \end{equation}
and

\begin{equation} D^{46.5}(r) = D_0 \exp [ - \beta^2 \mbox{4.07} (1-r) +\beta \mbox{0.46} (1-r^2) ] \ .\end{equation} 
When numerical calculations are evaluated at the reduced folding temperature of protein G  ($T_F=0.41$, \cite{Brown04}), equation (\ref{gq7}) with $G^{56}(r)$ and $D^{56}(r)$ predicts a mean folding time of $\tau_F \approx 10^{-1}\mbox{s}$. Likewise, $\tau_F \approx 10^{-2}\mbox{s}$ when taking $G^{46.5}(r)$ and $D^{46.5}(r)$. Thus, $G^{51}(r)$ is a suitable effective free energy landscape for the case when the last 10 codons are translated at the same average rate $\langle \tau_A \rangle$ and $G^{46.5}(r)$ is appropriate for the case when a synonymous codon mutation with a new average dwell time of $10 \times \langle \tau_A \rangle$ is made at position 81. This confirms that the effect of variable translation rates on folding times and pathways can be described by changes in an effective free-energy landscape whose shape reflects the highly complex dynamics of the underlying system. Furthermore, that this landscape can be derived directly from experimental observables. 
              
\section{Conclusions}

In this paper, two key concepts have been introduced for an energy landscape theory of cotranslational folding. The first is to consider each polypeptide of nascent chain length $n$ as a protein in its own right with a folding trajectory that is projected onto an $N$-dimensional nested energy landscape. A formal definition reveals that the ability of a protein to attain a stable fold at length $n$ is completely determined by curvatures of these landscapes. Secondly, cotranslational folding that is under kinetic control can be described by a diffusion process on an effective energy landscape: a weighted average of nested energy landscapes whose weights depend on the translation rate of individual codons. These concepts can be used to extend the current understanding of how cotranslational folders are designed to accommodate variable translation rates.       

Large fluctuations in average curvatures have previously been demonstrated to mark folding transitions and distinguish good folders from arbitrary polymer sequences \cite{Mazzoni06,Mazzoni08}. The origin of this behaviour is not entirely clear for the case of a full-length protein, but was originally suggested to be a result of the effective two-state dynamics of folding being connected to a symmetry-breaking mechanism of finite phase transitions \cite{Caiani98}. In the context of cotranslational folding, the average curvature of each nested energy landscape has a lower bound so that folding trajectories can converge on the ribosome. The conditions on the landscape at which convergence occurs have now been derived, and the development of experiments to study proteins arrested on the ribosome at different chain lengths \cite{Kaiser11} will be useful for confirming these results.        

Considering cotranslational folding as a composite stochastic process generalises two-state folding models with exponentially distributed dwell times at each codon \cite{O'Brien12}. This allows the Bryngelson and Wolynes approximation to make accurate predictions of mean folding times, or derive a suitable energy landscape for the entire reaction process from a set of experimentally observable parameters. The changes in the effective free energy landscape and diffusion coefficient can provide a good indication of how folding pathways and yields become altered by variable translation rates when the process in under kinetic control. Moreover, the theory can be used in combination with previous findings \cite{Ciryam13} to study the extent of cotranslational folding on a genome-wide scale since it provides an intuitive understanding of how proteins evolved to fold on the ribosome.   

% If in two-column mode, this environment will change to single-column
% format so that long equations can be displayed. Use
%\begin{widetext}
%\end{widetext}

% figures should be put into the text as floats.
% Use the graphics or graphicx packages (distributed with LaTeX2e)
% and the \includegraphics macro defined in those packages.
% See the LaTeX Graphics Companion by Michel Goosens, Sebastian Rahtz,
% entire page. There is no need to do explicit centering.

% \begin{figure}
% \end{figure}

% Surround figure environment with turnpage environment for landscape
% figure
%\begin{turnpage}

%\end{turnpage}

% tables should appear as floats within the text
% The ruledtabular enviroment adds doubled rules to table and sets a
% reasonable default table settings.
% Use the table* environment to get a full-width table in two-column
% Add \usepackage{longtable} and the longtable (or longtable*}
% environment for nicely formatted long tables. Or use the the [H]
% placement option to break a long table (with less control than 
% \begin{table}%[H] add [H] placement to break table across pages
% \begin{ruledtabular}
% Lines of table here ending with \\
% \end{tabular}
% \end{ruledtabular}
% \end{table}

% Surround table environment with turnpage environment for landscape
% table
% \begin{table}
% \begin{ruledtabular}
% \end{tabular}
% \end{ruledtabular}
% \end{table}

%\appendix
%\section{}

\begin{acknowledgments}
This work was supported by the Medical Research Council MC U105184332.
\end{acknowledgments}

% \bibliography{apssamp}

\end{document}